\documentclass[twocolumn,aps,superscriptaddress,showpacs,nofootinbib,floatfix]{revtex4-1}
\usepackage{epsfig,bm,feynmf}
\usepackage{graphics}
\usepackage{amsmath}
\usepackage[eps]{pstricks}
\usepackage{physics}
\usepackage{slashed}
\usepackage{dsfont}
\usepackage{multirow}
\usepackage {hyperref}
\usepackage[normalem]{ulem}  

\renewcommand\sout{\bgroup \color{red} \ULdepth=-.5ex \ULset}

\newcommand{\als}{\alpha_{s}}

\newcommand{\gev}{\text{GeV}}

\newcommand{\gm}[1][\mu]{\gamma_#1}
\newcommand{\ld}[1][a]{\lambda^#1}

\begin{document}


\title{The masses of hadrons in the chiral symmetry restored vacuum}


\author{Jisu Kim}%
\email{fermion0514@yonsei.ac.kr}
\affiliation{Department of Physics and Institute of Physics and Applied Physics, Yonsei University, Seoul 03722, Korea}

\author{Su Houng Lee}%
\email{suhoung@yonsei.ac.kr}
\affiliation{Department of Physics and Institute of Physics and Applied Physics, Yonsei University, Seoul 03722, Korea}


\begin{abstract}
We calculate the masses of the vector and axial vector mesons as well as the nucleon and the delta resonance in the chiral symmetry restored vacuum. This is accomplished by separating the  quark operators appearing in the QCD sum rules for these hadrons into the chiral symmetric and symmetry breaking parts depending on the contributions of the fermion zero modes.
We then extract the vacuum expectation values of all the separated parts of the quark operators using the QCD sum rule relations for these hadrons with their vacuum masses and widths.  
 By taking the chiral symmetry breaking parts to be zero while keeping the symmetric operators to their vacuum values, we obtain the chiral symmetric part of the hadron masses. 
  We find that the masses of chiral partners, such as the $(\rho,a_1)$ and $(K^*,K_1)$, become degenerate to values between 500 and 600 MeV in the chiral symmetry restored vacuum, while parity partners $(\omega,f_1)$ that are chiral partners only in the limit where the disconnected diagrams are neglected remain non-degenerate with masses $(655,1060)$ MeV, respectively.  
  The masses of the nucleon and the Delta are also found to reduce to about 500 and 600 MeV, respectively, in the chiral symmetric vacuum.  This shows 
that while   chiral symmetry breaking is responsible for the mass difference between chiral partner, both the meson and baryon retain non-trivial fraction of their masses in the chiral symmetry restored vacuum.  
\end{abstract}


\maketitle

\section{Introduction}

Understanding the generation  of the masses of hadrons that are larger than several hundred MeV starting from the current quark masses of less than 10 MeV is one of the fundamental problems in QCD\cite{Wilczek:1999be,Wilczek:2012sb}.  
It is believed that spontaneous chiral symmetry breaking\cite{Nambu:1961tp,Nambu:1961fr} is partly responsible for the generation of the masses \cite{Hatsuda:1985eb,Brown:1991kk,Hatsuda:1991ez,Leupold:2009kz}.  

Experiments have been performed worldwide to observe mass shift of hadrons at finite temperature or density \cite{Hayano:2008vn,Metag:2017yuh,Ichikawa:2018woh,Ohnishi:2019cif,Salabura:2020tou}, as chiral symmetry is expected to be fully and partially restored in the initial states of the relativistic heavy ion collision and in the nuclear matter probed by nuclear target experiment, respectively. 

It is clear that chiral symmetry breaking is responsible for the mass difference between chiral partners\cite{Weinberg:1967kj}, so that if chiral symmetry is restored, the mass difference between chiral partners should vanish\cite{Song:2018plu,Lee:2019tvt}. 
However, how much of the total hadron mass comes from chiral symmetry breaking is still to be understood.  

Recently, we have shown that the masses of the $\rho$ and $a_1$ mesons, which form chiral partners, become degenerate to about 550 to  600 MeV in the chiral symmetry restored vacuum\cite{Kim:2020zae}.  One can can isolate chiral symmetry breaking effect  by  separating the quark operators, appearing in the operator product expansion (OPE) of their correlation functions, into the chiral symmetric and breaking parts through their dependencies on the fermion zero mode, which can be directly related to the chiral order parameter\cite{BC}.  
As the four and six-quark operators are separated according to the contribution of the zero modes, one notes that the chiral symmetry breaking parts are order parameters of chiral symmetry breaking while the symmetric parts obtain non-vanishing values from chiral symmetric  non-perturbative contributions in QCD\cite{Kim:2020zae}.   
Originally, in the  in-medium QCD sum rules for the light vector mesons\cite{Hatsuda:1991ez}, 
the  four-quark condensates were assumed to be proportional to the quark condensate square through the vacuum saturation hypothesis\cite{Shifman:1978bx} so that they vanished automatically when chiral symmetry is restored.   
However, it was found that the chiral symmetric part of the four-quark condensate is  as large as the chiral symmetry breaking part and that the magnitude of the  chiral symmetry breaking part follows that obtained from using the vacuum saturation hypothesis\cite{Kim:2020zae}.  This also explains why in the previous QCD sum rules, the four-quark operators were multiplied by a $\kappa$ factor larger than 1 after vacuum saturation hypothesis to correctly obtain the hadron mass from the sum rule analysis\cite{Leupold:1998bt,Leupold:2001hj}.

In this work, we generalize the the work on ($\rho,a_1$)\cite{Kim:2020zae} to study the masses of other chiral partners ($K^{*}(892), K_{1}(1270)$) and parity partners ($\omega(782)$, $f_{1}(1285)$) as well as the  nucleon and the Delta resonance  in the chiral symmetry restored vacuum.  In section II, we lay out the formalism and the analysis method.  In section III, we apply the formalism to the meson sector. The baryons are analyzed in section IV.   Section V summarizes our work.  Finally, in the appendix, we provide detailed explanation of how to separate the quark operators into the chiral symmetric and breaking parts.  We further provide explicit forms for the chiral symmetric and breaking quark operators appearing in the meson and baryon sum rules and show relevant relations based on Fiertz transformations.  

\section{Sum rules analysis}
In this section, we provide an overview of the analysis method used in this work. We start with the relevant correlation function between  currents with the quantum numbers of the hadrons to study.  
\begin{eqnarray}
\Pi(q) & = & i \int d^4x \;e^{iqx} \langle T\{j^\Gamma(x) J^\Gamma(0)\} \rangle,
\end{eqnarray}
where the subscript $\Gamma$ specifies the current of interest. 
We then study the Borel transformed dispersion relation for the invariant part of the correlation function.  
\begin{eqnarray}
\widehat{\Pi}(M^2) = \int_{0}^{\infty} ds \;e^{-s/M^2} \rho(s),
\label{sum_rules}
\end{eqnarray}
where $\widehat{\Pi}(M^2)$ represents the Borel transformed OPE  of the correlator $\Pi$, and $M$ stands for the Borel mass.  $\rho(s)$ is the spectral function for the hadron of interest.

\subsection{Phenomenological Side}
The phenomenological side is constructed using the following form of the spectral density function $\rho(s)$.
 \begin{equation}
 \begin{split}
 \rho(s) &= \rho^{\text{pole}}(s)	 + \rho^{\text{cont}}(s), \\
\rho^{\mathrm{pole}}(s) &= \frac{1}{\pi}
\frac{f \Gamma \sqrt{s}}{(s - m^2)^2 + s\Gamma^2},  \label{pheno_side_pole}
\\
\rho^{\mathrm{cont}}(s) &= \frac{1}{\pi} \theta(s - s_0) \mathrm{Im} \tilde{\Pi}^{\mathrm{pert}}(s),
\end{split}
\end{equation}
where $m$($\Gamma$) is the Breit-Wigner mass(width) of a hadron and $s_{0}$ is the threshold parameter. The threshold parameter is chosen to  make the Borel curve flattest.
In table~\ref{tab1}, we summarize the masses and widths of hadrons we study and the corresponding interpolating currents used.  
These parameters with the spectral density will be used in Eq. \eqref{sum_rules} to obtain the values of the four quark operators appearing  in the operator product expansion (OPE) up to dimension-6 operators.
In a previous work, using such a method, we were able to obtain the chiral symmetry breaking and symmetric operators appearing in the $\rho-a_1$ sum rules\cite{Kim:2020zae}.  Here, we will generalize the method to other hadrons.
%
%
%
%
\begingroup
\renewcommand{\arraystretch}{1.5}
\begin{center}
\begin{table}[htbp]
\begin{tabular}{c | c c l} \hline
	\hline
Particle & Mass(MeV) & Width(MeV) & Interpolating Current\\
\hline
$\rho$    & 775.26   & 149.1   & $\bar{q} \tau^{3}\gm q$\\
$a_{1}$   &  1230 & 425   & $\eta_{\mu \nu}\bar{q} \tau^{3} \gm[\nu] \gm[5] q$\\
$\omega$  & 782.65 & 8.49   & $\bar{q} \gm q$\\
$f_{1}$   & 1281.9  & 22.7   & $\eta_{\mu \nu}\bar{q} \gm[\nu] \gm[5] q$\\
$K^{*} $  & 895.81 & 47.4  & $\bar{u} \gm s$\\
$K_{1}$   & 1272  & 90    & $\bar{u}\gm \gm[5] s$\\ 
$N$       & 938   & 0     &$\epsilon_{abc}(u_{a}C\gm u_{b})\gm[5]\gm d_{c}$\\
$\Delta$  & 1232  & 117   &$\epsilon_{abc}(u_{a}C\gm u_{b}) u_{c}$\\
	\hline 
\end{tabular}
\caption{The mass and width for particles used in Eq.~(\ref{pheno_side_pole}). The last column shows the interpolating current used for each particle. $q$ stands for the isospin doublet and $\eta_{\mu \nu}= q_{\mu} q_{\nu}/q^{2} - g_{\mu \nu}$. }
\label{tab1}
\end{table}
\end{center}
\endgroup
\subsection{OPE Side}
%
%
%
%
In the original QCD sum rules for light hadrons, the OPE were calculated up to dimension-6 four quark operators.  Then the vacuum saturation approximations were used to estimate the vacuum expectation values for these four-quark operators. The total OPE were then  used in the Borel transformed sum rules to estimate the mass of the particle.  
Unfortunately, factorization assumption is valid only in the  large $N_{c}$ limit and  in fact a multiplicative $\kappa$ were often introduce to better reproduce the hadron mass.   
On the other hand, if the phenomenological side is modelled  with physical observables, one can use the sum rule to estimate the values of the four-quark operators.  
The four-quark operators can then be separated into chiral symmetry breaking and symmetric parts depending on the contribution of the zero modes. 
 
According to the Banks-Casher formula\cite{BC,Cohen:1996ng}, the chiral order parameter is  proportional to the density of zero eigenvalues in the Euclidean formalism.
\begin{equation} 
\begin{split} 
\langle \bar{q} q \rangle &  =   \lim_{x \rightarrow 0} 
- \frac{1}{2} \langle {\rm Tr} [ S(0,x)- i\gamma_5  S(0,x) i \gamma_5 ] \rangle \\ 
&  =  -\pi \langle \rho(\lambda=0) \rangle.
\end{split} \label{CB-formula}
\end{equation}
In a previous work, we showed that the four-quark operator can also be divided according to the contribution of these zero modes\cite{Lee:2019tvt,Kim:2020zae}.  The  chiral symmetry breaking four quark operators  are proportional to the zero modes while symmetric operator has no contribution from them. In other words, while the former can be considered as chiral order parameters, the latter originates from other vacuum structure associated with non-zero eigenvalues. Therefore, when the zero mode contributions are taken away,  chiral symmetry breaking is restored and the breaking operators vanish while the symmetric operators remain the same.

Using our prescription for dividing the four quark operators, one can determine the chiral symmetry breaking part and chiral symmetric part of the four quark operators appearing as two independent combinations in  the sum rules of  chiral partners, such as the  $\rho,a_1$, and then use the two sum rules to determine the two values separately.  This allows us to estimate the mass of of vector meson  mass  by taking the chiral symmetry breaking operator to zero while keeping the chiral symmetric four quark operator to its vacuum value and studying the sum rule for the common $\rho,a_1$ mass.

The situation is more complicated for the $K^*, K_1$ or the $\omega, f_1$ cases.
For $K^*,K_1$,  they form SU(2) chiral partners, so that the chiral symmetry breaking operator proportional to both the strange and light quark condensate can be identified unambiguously.  Still, to obtain their mass in the chiral symmetry restored vacuum, one should use the information about the chiral symmetry breaking operator obtained in the $\rho, a_1$ sum rules. The mass at the chiral symmetry restored vacuum can then be estimated by taking all the chiral symmetry breaking operators to zero.   
As for the $\omega, f_1$, they form chiral partners only when contributions from disconnected diagrams are neglected.  As we will see, combining the results from the $\rho, a_1$ sum rule, we can estimate the magnitude of these operators and estimate the contribution of chiral symmetry breaking effects on  the $\omega, f_1$ masses.  For the nucleon and delta masses, we will use results from the  meson sum rules on the four quark operators and try to extract the values for the dimension-8 operators.
 
 A way to parametrize the the sum of the chiral symmetry breaking and symmetric four quark operator appearing in each sum rules is by introducing an auxiliary parameter $\kappa$ multiplying the vacuum saturation value of the chiral breaking operators so that a $\kappa$ value close to 1 means small contribution from the chiral symmetric operators. 

 The input values for the low dimensional OPE terms are given in Table \ref{tab2}. We use the quark condensate value evaluated from the GMOR relation,
\begin{equation} 
\begin{split} 
m_{\pi}^{2}f_{\pi}^{2} = -2 m_{q} \ev{\bar{u}u},
\end{split} 
\end{equation}  
with the values ($m_{\pi}$ = 137.5 MeV, $f_\pi$ = 93 MeV, $\bar{m}_{ud}$ = 3.45 $\times$ 1.35 MeV). $m_{\pi}$ and $f_{\pi}$ are the mass and decay constant of the pion respectively. $m_{q}$ is the averaged mass of $u$ and $d$ quarks. We take the values of $m_{q}$ and $m_{s}$ scaled from 2 GeV to 1 GeV, as reported in Particle Data Group\cite{ParticleDataGroup:2020ssz}, implemented here through the factor of 1.35 multiplying the quark mass. Also, we take the values of the other sum rule parameters to be values that are frequently used in QCD sum rule studies. The other parameters are standard values taken from \cite{Hatsuda:1992bv,Ioffe:1983ju}
\begin{center}
\begingroup
\setlength{\tabcolsep}{12pt} 
\renewcommand{\arraystretch}{1.5} 
\begin{table}[htbp]
\begin{tabular}{l  l} \hline
	\hline
	$m_{q}$ & 3.45 $\times$ 1.35 MeV\\
	$m_{s}$ & 93 $\times$ 1.35 MeV\\
$\ev{\bar{u}u}$ & (0.260 GeV)$^3$\\
$\ev{(\als/\pi) G^{2}}$ & 0.012 GeV$^4$\\
$\als$ & 0.36 \\
 $\ev{\bar{s}s}/\ev{\bar{u}u}$ & 0.8\\
 $m_{0}^{2}$ & 0.8 $\mathrm{GeV}^{2}$\\
 $\mu$ & 0.5 GeV\\
 $\Lambda_{\mathrm{QCD}}$ & 0.1 GeV\\ \hline
 $ \ev{B_{su}}_{B}$/$\ev{B_{uu}}_{B}$ & 0.565(*)\\
  $\ev{B_{uu}}_{B}$ & $(0.3053 \;\mathrm{GeV})^{6}$(*)\\
  $\ev{B_{su}}_{B}$ & $(0.2776 \; \mathrm{GeV})^6$(*)\\
$\ev{(\bar{q}\gm \gm[5] \ld q)^{2}}_{dis,S}$ &  $(0.2826\; \mathrm{GeV})^{6}$(*)\\
$\ev{(\bar{q}\gm \ld q)^{2}}_{dis,S}$ &  $-(0.3369 \; \mathrm{GeV})^{6}$(*)
	\\ \hline 
\end{tabular}
\caption{The upper part shows the  OPE parameters used in the sum rule.  Lower part marked with a star marker(*) shows values calculated from the sum rules. $m^{2}_{0}$ stands for $g_{s}\ev{\bar{u} \sigma_{\mu \nu}G^{\mu \nu}_{a} \lambda^{a}u}/\ev{\bar{u}u}$. All the values here are obtained with assuming $a_1$ width to be 425 MeV. }
\label{tab2}
\end{table}
\end{center}
\endgroup

\subsection{Borel Window}

We now discuss how the Borel windows are obtained in the sum rules.  
The extracted sum rule is obtained by setting the phenomenological side $\widehat{\Pi}^{\mathrm{phen.}}(M^{2})$ equal to the OPE side $\widehat{\Pi}^{\mathrm{OPE}}(M^{2})$. Each side can be defined as follows.
\begin{equation} 
\begin{split} 
\widehat{\Pi}^{\mathrm{phen.}}(M^{2}) =& \int^{\infty}_{0} ds\; e^{-s/M^{2}} (\rho^{\mathrm{pole}}(s) +\rho^{\mathrm{cont}}(s)),\\
\widehat{\Pi}^{\mathrm{OPE}}(M^{2}) =& \sum_{j = d_{l}}^{d_{h}} \frac{C_{j}\ev{O_{j}}}{(M^{2})^{j}},
\end{split} 
\end{equation}
where $O_{j}$ and $C_{j}$ are the operator and the corresponding Wilson coefficient, respectively, and $d_{l}$($d_{h}$) is the lowest(highest) power of $(M^{2})^{-1}$ considered. 
As we truncate the OPE, the asymptotic expansion will break down at small Borel mass region.  On the other hand, the approximation of the continuum approximation and the pole dominance   require the Borel mass to sufficiently small.   The constraints give the acceptable Borel mass range called Borel window. The details are as follows.

In the phenomenological side, one has to make sure that the contribution from the pole $\rho^{\mathrm{pole}}(s)$ dominates over that from the continuum $\rho^{\mathrm{cont}}(s)$.   The constraint is given by
\begin{equation} 
\begin{split} 
\frac{\int_{s_0}^{\infty}ds\; e^{-s/M^{2}}\rho^{\mathrm{cont}}(s)}{\int_{0}^{\infty}ds\; e^{-s/M^{2}}\rho^{\mathrm{pole}}(s)} < x_{\mathrm{max}},
\label{bm_max}
\end{split} 
\end{equation}
where $x_{\mathrm{max}}$ is a number  smaller than 1 and adjusted for the sum rule for each hadron. This condition determines the upper boundary $M_{\mathrm{max}}$ in the Borel window. 

The other comes from requiring the contribution of the condensate terms to be smaller than that of the perturbative term. This constraint is obtained by
\begin{equation} 
\begin{split} 
\frac{\widehat{\Pi}^{\mathrm{OPE}}_{\mathrm{cond\; terms}}(M^{2})}{\widehat{\Pi}^{\mathrm{OPE}}_{\mathrm{pert\; term}}(M^{2})} < x_{\mathrm{min}},
\label{bm_min}
\end{split} 
\end{equation}
where $\widehat{\Pi}^{\mathrm{OPE}}_{\mathrm{cond\; terms}}(M^{2})$ and $\widehat{\Pi}^{\mathrm{OPE}}_{\mathrm{pert\; term}}(M^{2})$ are the sum of the condensate terms considered and the perturbative term in OPE side, respectively. This condition restricts the Borel mass $M$ to be larger than $M_{\mathrm{min}}$.

The values of $x_{\mathrm{max}}$ and $x_{\mathrm{min}}$ are chosen to make the extreme point of the Borel curve stable against changes in the threshold within the Borel window. The chosen values for different hadrons are given in Table~\ref{bwx}, which will also be used in the sum rule analysis in the chiral symmetry restored vacuum.
With these, the ranges of Borel window for each hadron can be determined. 
Once the Borel range is determined for a given hadron, the same range will be used to determine the mass in the chiral symmetry restored vacuum.
 In the following Borel curves presented in this work, Borel curves within(outside) the Borel window are drawn in a solid(dotted) line. 

\subsection{Analysis}

To eliminate the $f$-dependence in $\rho^{\mathrm{pole}}(s)$ shown in  Eq~(\ref{pheno_side_pole}), we conduct an analysis using the ratio between the extracted sum rule and its derivative with respect to $-1/M^{2}$. Namely,
\begin{equation} 
\begin{split} 
\frac{\frac{d}{d(-1/M^{2})}\widehat{\Pi}^{\mathrm{phen.}}(M^{2})}{\widehat{\Pi}^{\mathrm{phen.}}(M^{2})}=\frac{\int^{\infty}_0 ds\;s e^{-s/M^{2}}  \rho(s)}{\int^{\infty}_0 ds\; e^{-s/M^{2}} \rho(s)}.
\label{phen_der}
\end{split} 
\end{equation}
Apart from the Borel mass $M$ dependence, Eq~(\ref{phen_der}) depends on the hadron mass $m$, the decay width $\Gamma$ and the threshold parameter $s_{0}$. Meanwhile, its corresponding OPE side
\begin{equation} 
\begin{split} 
\frac{\frac{d}{d(-1/M^{2})}\widehat{\Pi}^{\mathrm{OPE}}(M^{2})}{\widehat{\Pi}^{\mathrm{OPE}}(M^{2})} = \frac{-\sum_{j=d_{l}}^{d_{h}}jC_{j}\ev{O_{j}}/(M^{2})^{j-1}}{\sum_{j=d_l}^{d_{h}}C_{j}\ev{O_j}/(M^{2})^{j}}
\label{OPE_der}
\end{split} 
\end{equation}
depends on the auxiliary parameter $\kappa$ for its dimension-6 operator term. As explained previously, the mass and decay width values in Table~\ref{tab1} are used. 
Equating Eq. \eqref{phen_der} to Eq. \eqref{OPE_der}, we obtain the Borel curve for $\kappa$.  We then take its value at the extremum point. Since the $\kappa$ value at the extremum point varies with $s_{0}$ value, we take the value that reproduce the flattest and stable mass Borel curve. 
In summary, the Borel sum rule we will use throughout the paper is given by
\begin{equation} 
\begin{split} 
\frac{\int^{\infty}_0 ds\;s\; e^{-s/M^{2}}  \rho(s)}{\int^{\infty}_0 ds\; e^{-s/M^{2}} \rho(s)} = \frac{-\sum_{j=d_{l}}^{d_{h}}jC_{j}\ev{O_{j}}/(M^{2})^{j-1}}{\sum_{j=d_l}^{d_{h}}C_{j}\ev{O_j}/(M^{2})^{j}}.
\label{basic-sumrule}
\end{split} 
\end{equation}

Once the $\kappa$ value is determined, the chiral symmetry breaking and symmetric quark operators can be evaluated by combing the sum rules of chiral partners. By taking chiral symmetric contribution only, one can get the mass sum rule in chiral symmetry restored vacuum. By taking the $s_0$ value that produce the flattest and stable mass Borel curve, the mass can be extracted. more details are discussed in the following subsections for different hadrons. 

We assume that in the symmetry restored phase, the decay widths for baryon the Delta, $\omega$ and $f_{1}$ remain the same, while for the chiral partners $K^{*}$-$K_{1}$ and $\rho$-$a_{1}$, they  become  that of the corresponding vector mesons. Our previous work\cite{Kim:2020zae} shows that changing the width  in the chiral symmetric vacuum, affects the symmetric mass $m_{sym}^{\rho-a_1}$ insignificantly, at most 20 MeV. This amount is less than 10\% of the total mass decrease. Furthermore, since the widths of other hadrons we considered are smaller  than that of $\rho$, the effects coming from using different widths are also expected to be small. 
\begingroup
\renewcommand{\arraystretch}{1.5}
\begin{center}
\begin{table}[htbp]
\begin{tabular}{c | c c c c c c c c} \hline
	\hline
 &$\rho$  & $a_{1}$ & $\omega$ & $f_{1}$ & $K^{*}$ & $K_{1}$ & $N$ & $\Delta$ \\
\hline
$x_{\mathrm{min}}$   & 0.15 & 0.15 & 0.15 & 0.15 & 0.15 & 0.15 & 0.2 & 0.28   \\
$x_{\mathrm{max}}$   & 0.7 & 0.7 & 0.7 & 0.7 & 0.7 & 0.7 & 0.75 & 0.96  \\
	\hline 
\end{tabular}
\caption{ Values for  $x_{\mathrm{max}}$ and $x_{\mathrm{min}}$ defined in Eq~(\ref{bm_max}) and (\ref{bm_min})}
\label{bwx}
\end{table}
\end{center}
\endgroup

\section {Meson}
While both ($\rho$, $a_1$) and ($K_{1}$, $K^{*}$) are chiral partners, ($\omega$, $f_{1}$) are not because of the contributions from the disconnected diagrams. 
These contributions are not zero as $\omega$($f_{1}$) and $\rho$($a_{1}$) have different widths suggesting that the differences in the OPE coming from the disconnected four quark contributions, which vanishes in the vacuum saturation hypothesis, are not zero.  Let us first summarize the four quark condensate appearing in the $\rho,a_1$ sum rules.  
%
%
%
%
\begin{center}
\begin{table}[htbp]
\begin{tabular}{c | c c c } \hline
	\hline
Particle & $\kappa(\sqrt{s_0}(\gev))$ & $S/B$ & $\bar{m}_{sym}$(MeV)\\
\hline
$\rho$  & 2.60(1.17) & 0.760 & \multirow{2}{*}{572.5 $\pm$ 27.5}\\
$a_{1}$ & 0.76(1.58) & -0.485 \\ \hline
$\omega$& 3.20(1.16) & 1.165 & 655 $\pm$ 15\\
$f_{1}$ & 1.85(1.58) & 0.253 & 1060 $\pm$ 30\\ \hline
$K^{*} $& 2.097(1.33) & 2.831&  \multirow{2}{*}{545 $\mp$ 5}\\
$K_{1}$ & 0.39(1.56) & -0.227&  \\ 
	\hline 
\end{tabular}
\caption{The $\kappa$'s are evaluated using sum rules with  the physical values given in Table~\ref{tab1} and the threshold parameter $s_{0}$ in the brackets.  $S/B$ indicates the fractional contribution  of the chiral symmetric part $S$ to the contribution of the chiral symmetry breaking part $B$ for each hadron. $\bar{m}_{sym}$ is the hadronic mass in the chiral symmetry restored vacuum. The uncertainties of the masses are due to the ranges of the width of $a_{1}$ meson, which we take from 250 MeV to 600 MeV, for the negative and positive uncertainty values, respectively. Except for the last column, the central  values are evaluated with the $a_{1}$ width of 425 MeV.}
\label{tab3}
\end{table}
\end{center}
%
%
%
%
%
\subsection{$\rho$ and $a_{1}$}

The sum rules are obtained by using the interpolating currents given in Table \ref{tab1} and studying the polarization $\Pi=\Pi^\mu_\mu/(-3q^2)$.  
The dimension-6 four-quark operators from the OPE sides contributing as 
$\mathcal{M}/Q^6$ 
 are respectively given as follows\cite{Hatsuda:1992bv}. 
\begin{equation} 
\begin{split}
 \mathcal{M}_{\rho} =&\; -2\pi \als\ev{(\bar{q} \gm \gm[5] \ld \tau^{3}q)^{2}} \\& - \frac{4\pi \als}{9}\ev{(\bar{q}\gm \ld q)(\sum_{q = u,d,s}\bar{q}\gm \ld q)}, \\
  \mathcal{M}_{a_{1}} =&\;  - 2\pi \als \ev{(\bar{q} \gm \ld \tau^{3}q)^{2}} \\& - \frac{4\pi \als}{9}\ev{(\bar{q}\gm \ld q)(\sum_{q = u,d,s}\bar{q}\gm \ld q)},
\end{split}
\end{equation} 
where $q$ denotes the isospin doublet unless stated otherwise. As mentioned in the previous section, these can be divided into the chiral symmetric and breaking operators; details are given in Appendix A. By omitting the common factor ($-\pi \als $) and introducing the auxiliary parameters $\kappa_{\rho}$ and $\kappa_{a_{1}}$ for each, the results are obtained as the below. 
\begin{equation} 
\begin{split} 
\kappa_{\rho}\frac{448}{81}\ev{\bar{u}u}^{2}=&\; \frac{28}{9}\ev{B_{uu}}_{B}+ \ev{S_{\rho-a_{1}}}_{S},\\
-\kappa_{a_{1}}\frac{704}{81}\ev{\bar{u}u}^{2}=&\; -\frac{44}{9}\ev{B_{uu}}_{B}+ \ev{S_{\rho-a_{1}}}_{S},
\end{split} 
\end{equation}
where 
\begin{equation} 
\begin{split} 
\ev{B_{uu}}_{B} =& \frac{1}{2}\bigg(\ev{(\bar{u}\gm \gm[5] \ld d)(\bar{d} \gm \gm[5] \ld u)}\\ 
&\;\;\;\;\;- \ev{(\bar{u}\gm \ld d)(\bar{d} \gm \ld u)}\bigg)_{B} , \\
\ev{S_{\rho-a_1}}_{S}  =& \frac{11}{9}\bigg(\ev{(\bar{q} \gm \gm[5] \ld \tau^{3} q)^{2}}+\ev{(\bar{q} \gm \ld \tau^{3} q)^{2}}\bigg)_{S}\\
&+\frac{4}{9}\bigg( \ev{(\bar{q}\gm \ld q)^{2}}- \ev{(\bar{q}\gm \ld \tau^{3}q)^{2}}\bigg)_{S}\\
&+\frac{4}{9}\ev{(\bar{q} \gm \ld q)(\bar{s} \gm \ld s)}_{S}.
\end{split} 
\end{equation}
Since the symmetric part is identical while the  breaking operator $B_{uu}$  contribute with different coefficients, one can identify the breaking and symmetric part uniquely by optimizing both sum rules with physical quantities. By taking the breaking part of the operator to be zero while keeping the symmetric operator to its vacuum value, one can obtain the sum rules in the chiral symmetry restored vacuum.  The results for the matrix elements and the masses in the chiral symmetry restored vacuum are given in Table \ref{tab3}. 
The results for $\rho-a_1$ in Table \ref{tab3} are obtained with the $a_1$ width of 425  MeV.  
Since there are still large uncertainty in the $a_1$ width $\Gamma_{a_{1}}$, we take a larger value.  For $\Gamma_{a_{1}}= 250 (600)$  MeV, one finds that $m_{sym}^{\rho-a_{1}}$ decreases (increases) by  about 27.5 MeV.

\subsection{$\omega$ and $f_{1}$}

\begin{figure}[h]
\centerline{
\includegraphics[width=9 cm]{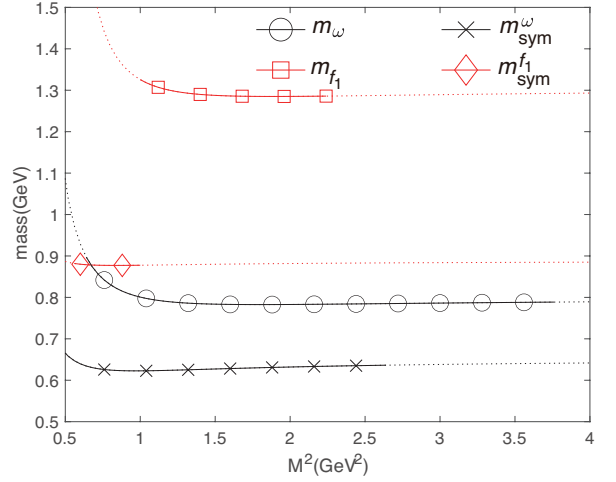}}
\caption{The Borel curves for $\omega$(open circle) and $f_{1}$(square) masses are with each $\kappa$ value in TABLE~\ref{tab3}. The Borel curve with cross(diamond) markers describes the $\omega$($f_{1}$) mass in the chiral symmetry restored vacuum}
\label{mass-wf}
\end{figure}

The only difference between the OPE sides of $\rho$($a_{1}$) and $\omega$($f_{1}$) is the disconnected piece\cite{Gubler:2016djf}. 
\begin{equation} 
\begin{split} 
\Pi^{\mathrm{OPE}}_{\omega} - \Pi^{\mathrm{OPE}}_{\rho} &= -\frac{2\pi\als}{Q^{6}}\ev{(\bar{q}\gm \gm[5] \ld q)^{2}}_{dis,S}, \\
\Pi^{\mathrm{OPE}}_{f_{1}} - \Pi^{\mathrm{OPE}}_{a_{1}} &= -\frac{2\pi \als }{Q^{6}}\ev{(\bar{q} \gm \ld q)}_{dis,S},
\label{difaf}
\end{split} 
\end{equation}
where the two-point correlators $ \Pi^{\mathrm{OPE}}_{\omega}$ and $\Pi^{\mathrm{OPE}}_{f_{1}}$ come from the interpolating currents $J_{\mu}^{\omega} $ and $J_{\mu}^{f_{1}} $ in Table~\ref{tab1}.  These disconnected 
operators are chiral symmetric and are mainly responsible for the differences in their masses and widths. As the chiral symmetry operators do not depend on chiral symmetry breaking effects, the masses of  $\omega$ and $f_{1}$ will be non-degenerate in the chiral symmetry restored vacuum. With the auxiliary parameter $\kappa$, Eq.~(\ref{difaf}), apart from (-$\pi \alpha_{s}/Q^{6}$) can be rewritten as below.
\begin{equation}
\begin{split}
	(\kappa_{\omega} - \kappa_{\rho})\frac{448}{81}\ev{\bar{u}u}^{2} &=2\ev{(\bar{q}\gm \gm[5] \ld q)^{2}}_{dis,S},\\
	-(\kappa_{f_{1}} - \kappa_{a_{1}})\frac{704}{81} \ev{\bar{u}u}^{2}&=2\ev{(\bar{q} \gm \ld q)}_{dis,S}.
	\label{disc_sym}
\end{split}	
\end{equation}
\begin{figure}[h]
\centerline{
\includegraphics[width=9 cm]{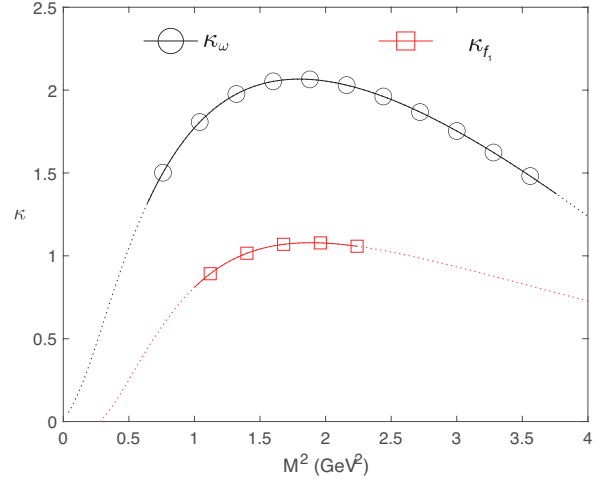}}
\caption{Borel curves for four-quark operators parametrized by $\kappa$ for $\omega$(circle) and $f_{1}$(square). The optimal $\kappa$ is taken to be the value at the extremum.}
\label{kappa-wf}
\end{figure}
The $\kappa$ values  can be obtained from the values for the four-quark operators, which are calculated by  substituting the phenomenological side given in Eq.~\eqref{pheno_side_pole} into Eq.~\eqref{sum_rules} and studying the sum rule for each current.  
$\kappa_\rho$ and $\kappa_{a_{1}}$ given in Table~\ref{tab3}  are obtained in the sum rule calculation presented in~\cite{Kim:2020zae}. Here the values for $\kappa_\rho$ and $\kappa_{a_{1}}$ are slightly different from those given in ~\cite{Kim:2020zae} as we use  updated values for the quark condensate and mass.
   Substituting the spectral density with physical  parameters for $\omega$ and $f_{1}$ into the sum rules, we obtain the corresponding $\kappa$ values as given in Table~\ref{tab3}.  Fig.\ref{kappa-wf} shows the $\kappa$ values for $\omega$ and $f_1$ meson.  We take the $\kappa$ value at the extremum point of the Borel curve.  One can also use the Borel curve for the masses and try to extract the value for the  four-quark operators that best reproduces the $\omega$ and $f_1$ mass with their given width.  As can be seen in the open circle and square plot in Fig.~\ref{mass-wf} the extracted value for the four-quark operators gives the most stable Borel curve for the mass of $\omega$ and $f_1$ mass, respectively.  Using the  values of  $\kappa$, the values for the disconnected contribution of the four-quark condensates are given as below.
\begin{equation} 
\begin{split} 
\ev{(\bar{q} \gm \gm[5] \ld q)^{2}}_{dis,S} = 1.65 \ev{(\bar{u}u)}^{2},\\
\ev{(\bar{q} \gm \ld q)^{2}}_{dis,S} = -4.73 \ev{(\bar{u}u)}^{2}.
\label{disc_val}
\end{split} 
\end{equation}
The disconnected contributions vanish in the vacuum saturation hypothesis.  Our result in Eq.~\eqref{disc_val} shows that while such assumptions are approximately valid in some channels, they are largely violated in some other channels.   
Assuming that the disconnected diagrams are mediated by gluon fields, one notes $\ev{(\bar{q} \gm \gm[5] \ld q)^{2}}_{dis,S} /\ev{(\bar{q} \gm \ld q)^{2}}_{dis,S} \sim (\alpha_{s})^{2}$ suggesting that estimated values in Eq.~(\ref{disc_val}) are consistent with the $\alpha_s$ counting.

 The magnitude of dimension-6 part for $\omega$($f_{1}$) consists of 
the disconnected piece and the dimension-6 part of $\rho$($a_{1}$).
Now, using the identification discussed in Appendix A, one notes that the disconnected four-quark operators appearing in Eq.~\eqref{difaf} are chiral symmetric operators.  
Therefore, the mass of the $\omega$ and $f_1$ mesons in the chiral symmetry restored vacuum can be obtained by taking the chiral symmetry breaking operators, which appear in the $\rho$ and $a_1$ channels, to zero.  
 Then, as can be seen in Fig.\ref{mass-wf} through the Borel curve, one finds that when the symmetry is restored, the $\omega$ mass becomes  650 MeV, similar to the $\rho-a_{1}$ mass, while the $f_{1}$ mass is 1060 MeV.

\subsection{$K_{1}$ and $K^{*}$}
\begin{figure}[h]
\centerline{
\includegraphics[width=9 cm]{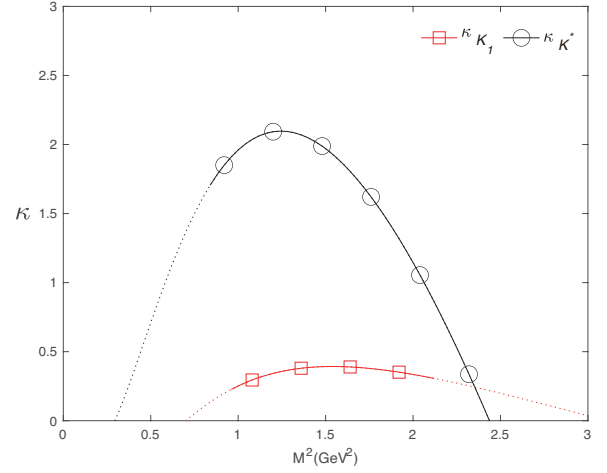}}
\caption{Borel curves for four-quark operators parametrized by $\kappa$ in $K^{*}$(circle) and $K_{1}$(square) sum rules. The optimal $\kappa$ values are taken from the value at the extremum point.}
\label{k1kp}
\end{figure}

\begin{figure}[h]
\centerline{
\includegraphics[width=9 cm]{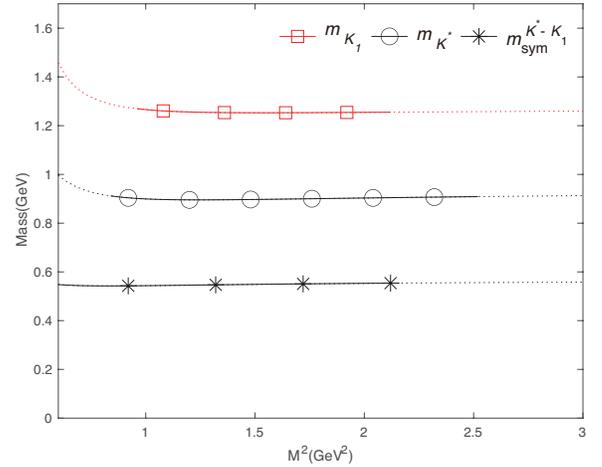}}
\caption{Borel curves for $K_{1}$ and $K^{*}$ mass in chiral symmetry breaking(square and circle) and restored(cross) vacuum. }
\label{k1mass}
\end{figure}

Unlike the $\omega, f_1$ pair, $K^{*}$ and $K_{1}$ form chiral partners, as is  evident when analyzing the  operators as given in Appendix A. Hence, when  the chiral symmetry gets restored, the masses become degenerate. As represented in \cite{Song:2018plu}, the dimension-6 operators, after omitting the common factor $-\frac{2\pi \alpha_{s}}{Q^{6}}$ in the OPE, are  given as below. 
\begin{equation} 
\begin{split} 
 \mathcal{M}_{K^{*}} =& \ev{( \bar{u}\gm \gm[5] \ld s )(\bar{s}\gm \gm[5] \ld u)} \\
&+ \frac{1}{9}\ev{(\bar{s} \gm \ld s+\bar{u}\gm \ld u)(\sum_{q=u,d,s} \bar{q} \gm \ld q)},\\
\mathcal{M}_{K_{1}} = & \ev{( \bar{u}\gm \ld s )(\bar{s}\gm \ld u)} \\
&+ \frac{1}{9}\ev{(\bar{s} \gm \ld s+\bar{u}\gm \ld u)(\sum_{q=u,d,s} \bar{q} \gm \ld q)}.
\end{split} 
\end{equation}
One can divide these operators according to chiral symmetric and breaking parts, shown by the subscripts $S$ and $B$,  respectively. 
\begin{equation} 
\begin{split} 
\mathcal{M}_{K^{*}} =&\ev{B_{su}}_{B}- \frac{1}{9}\ev{B_{uu}}_{B} +\ev{S_{K}}_{S}, \\
\mathcal{M}_{K_{1}} = &-\ev{B_{su}}_{B}-\frac{1}{9} \ev{B_{uu}}_{B} +\ev{S_{K}}_{S},
\label{k-seq}
\end{split} 
\end{equation}
where \begin{equation} 
\begin{split} 
\ev{B_{su}}_{B} &= \frac{1}{2}\ev{(\bar{u} \gm \gm[5] \ld s)(\bar{s} \gm \gm[5] \ld u)-(\bar{u} \gm  \ld s)(\bar{s} \gm  \ld u) }_{B},\\
\ev{B_{uu}}_{B} &= \frac{1}{2}\ev{(\bar{u} \gm \gm[5] \ld d)(\bar{d} \gm \gm[5] \ld u)-(\bar{u} \gm  \ld d)(\bar{d} \gm  \ld u) }_{B},\\
\ev{S_{K}}_{S} &= \frac{1}{2}\ev{(\bar{s} \gm \gm[5] \ld u)(\bar{u} \gm \gm[5] \ld s) +(\bar{s} \gm \ld u)(\bar{u} \gm \ld s) }_{S}\\
 +&\frac{1}{18}\ev{(\bar{u}\gm \gm[5] \ld d)(\bar{d} \gm \gm[5] \ld u)+(\bar{u} \gm \ld d)(\bar{d} \gm \ld u)}_{S}\\
+&\frac{1}{9}\bigg( \langle(\bar{s}\gm \ld s)(\sum_{q=uds}\bar{q} \gm \ld q)\rangle_{S} \\
&\;\;+ \ev{(\bar{u} \gm \ld u)(\bar{d} \gm \ld d)}_{S}+ \ev{(\bar{u}\gm \ld u)(\bar{s} \gm \ld s)}_{S}\\
&\;\;+\ev{(\bar{u} \gm \ld u)^{2}-(\bar{u} \gm \ld d)(\bar{d} \gm \ld u)}_{S} \bigg),
\label{bsbu}
\end{split} 
\end{equation}
and $S_{K}$ is the symmetric dimension-6 operators for $K^{*}$ and $K_{1}$.  Using the sum rules for the $K^*$ and $K_1$ mesons one can obtain the Borel curves and thus determine the  magnitudes of the four-quark operators appearing in these sum rules separately.  We parametrize the magnitude of the obtained matrix elements by identifying them to $\kappa_{K^*}$ and $\kappa_{K_1}$ times the  vacuum vacuum saturation values, respectively.   Eq.~(\ref{k-seq}) can be rewritten as follows. 

\begin{equation} 
\begin{split} 
 \mathcal{M}_{K^{*}} =& \kappa_{K^{*}}\bigg(\frac{16}{9}\ev{\bar{u}u}\ev{\bar{s}s} -\frac{16}{81}(\ev{\bar{u}u}^{2}+\ev{\bar{s}s}^{2})\bigg) \\
 =&\ev{B_{su}}_{B}-\frac{1}{9} \ev{B_{uu}}_{B} +\ev{S_{K}}_{S}, \\
\mathcal{M}_{K_{1}} =& \kappa_{K_{1}}\bigg(-\frac{16}{9}\ev{\bar{u}u}\ev{\bar{s}s} - \frac{16}{81}(\ev{\bar{u}u}^{2}+\ev{\bar{s}s}^{2})\bigg)\\
  = &-\ev{B_{su}}_{B}-\frac{1}{9}\ev{B_{uu}}_{B} +\ev{S_{K}}_{S}.
\label{kkk2}
\end{split} 
\end{equation}
Thus, the Borel curves for the four-quark operators become those for the $\kappa$'s. 
It should be noted that one can first determine $\ev{B_{su}}_{B}$ from the difference between the $K^*-K_1$ sum rules.  As can be seen from Table \ref{tab4}, the value of $\ev{B_{su}}_{B}$ thus determined is slightly smaller than $\ev{B_{uu}}$  determined from the  $\rho$-$a_{1}$ sum rules, which is what is expected also from the vacuum saturation hypothesis as the strange quark condensate is expected to be slightly smaller than that of light quark condensate.  Using these values, one can further  determine the values for $\ev{S_{K}}_{S}$.  
Fig.~\ref{k1kp} shows the Borel curve for the $\kappa$ values for the pair. As in the previous section, the values are taken at the extremum point and also the threshold parameter ($\sqrt{s})$ values are taken to make the mass Borel curve flattest and stable. One can use the obtained values for the four-quark operators and obtain the sum rule for the masses.  As presented in Fig. \ref{k1mass}, the  Borel curves for the masses of the $K^{*}$(open circle) and $K_{1}$(square) with the corresponding $\kappa$ value are stable and reproduce the physical masses.  

The chiral symmetric degenerate mass can be extracted by taking the symmetry breaking operator to be zero and keeping the magnitude of the symmetric operator $\ev{S_{K}}$. Although the $\ev{B_{uu}}$ value used to solve Eq.~(\ref{k-seq}) depends on the $a_{1}$ width, the extracted mass $m_{sym}^{K^{*}-K_{1}}$ barely changes as one changes  its value. One finds that   $m_{sym}^{K^{*}-K_{1}}$ is 545 MeV, slightly smaller than that of $\rho-a_{1}$. For $\Gamma_{a_{1}}= 250 (600)$  MeV, one finds that $m_{sym}^{K^{*}-K_{1}}$ increases (decreases) by  about 5 MeV. 

\section{Baryon}
We use the currents given in Table \ref{tab1}.  Since the baryon currents couple to both parity states,  
the two-point correlator between the  baryonic interpolating current $\eta(x)$ can be written as follows\cite{Gubler2013}. 
\begin{equation} 
\begin{split} 
\Pi(q) =&\; i\int d^{4}x\; e^{iqx} \ev{T\{\eta(x) \bar{\eta}(0)\}}{0}\\
=& \;\sum_{n} \bigg\{-\abs{\lambda^{n}_{+}}^{2} \frac{\slashed{q} +m^{n}_{+}}{q^{2} -(m_{+}^{n})^{2} +i\epsilon} \\
&\qquad \;\; \;- \abs{\lambda^{n}_{-}}^{2} \frac{\slashed{q} - m^{n}_{-}}{q^{2} - (m^{n}_{-})^{2} + i \epsilon}\bigg\} \\
\equiv &\; \slashed{q} \,\Pi_{v}(q^{2}) + \mathds{1}\,\Pi_{s}(q^{2}) ,
\end{split} 
\end{equation}
where $m_+^n$ and $m_-^n$ represent the mass of the $n$-th excited positive and negative parity states, respectively, and $\lambda_\pm^n$ their corresponding overlaps with the currents.  
As two invariant functions appear, one can obtain two independent sum rules. However, if chiral symmetry is restored, $\Pi_{s}(q^{2})$ goes to zero as  all the operators in the OPE side are chiral order parameters and the phenomenological side will be proportional to the mass difference between chiral partners. 
Therefore we will use the sum rule from $\Pi_{v}(q^{2})$, from which we can extract the mass even  in the chiral symmetry  restored phase as the two chiral partners will become degenerate and strengthen the lowest pole contribution. 

Since we do not have independent sum rules for chiral partners, we can not determine the values for the chiral symmetric and breaking quark operators separately as we did in the meson case.
On the other hand, as can be seen in Table \ref{tab4}, the values of the chiral symmetry breaking four-quark operator extracted from  the meson sum rule 
are very close to that estimated from the vacuum saturation hypothesis.  
 Therefore, we will assume that the vacuum saturation hypothesis is accurate for evaluating the chiral breaking operators, as was the case for four-quark operators appearing in the meson sum rules. 
\begingroup
\renewcommand{\arraystretch}{1.5}
\begin{center}
\begin{table}[htbp]
\begin{tabular}{c | c c c} \hline
	\hline
Operator & SR & VS & Error\\
\hline
$\ev{B_{uu}}_{B}$ & (0.3056 GeV)$^6$   & (0.2862 GeV)$^6$ & 0.482\\
$\ev{B_{su}}_{B}$ & (0.2779 GeV)$^6$ & (0.2758 GeV)$^6$ & 0.038\\
$\ev{(\bar{q} \gm \gm[5] \ld q)^{2}}_{dis,S}$ & (0.2826 GeV)$^6$   & 0 GeV$^6$ & \\
$\ev{(\bar{q} \gm \ld q)^{2}}_{dis,S}$   & $-$(0.3369 GeV)$^6$ & 0 GeV$^6$  & \\
\hline 
\end{tabular}
\caption{ Values of the four four-quark operators defined in Eq.~\eqref{bsbu}  evaluated from our sum rule approach (SR) and   from vacuum saturation hypothesis (VS).   The error shown in the last column are the difference between the two divided by VS value.}
\label{tab4}
\end{table}
\end{center}
\endgroup

Furthermore, while the OPE was calculated 
 up to dimension-6  four quark operators for the meson case, one needs to 
consider the higher dimensional  four quark gluon mixed operator for the baryons in order to obtain stability of the sum rules.   
We extract and parametrize the magnitude of the dimension-6 and dimension-8 operators by $\kappa_{6}$ and $\kappa_{8}$, respectively, as was done for the dimension-6 operators in the meson sum rules. 

In this case, there is one more independent parameter $\kappa_{8}$.
In principle, since the sum rule is a function of Borel mass, we can determine both parameters by requiring the phenomenological side to be closest to the OPE side.  Technically, we realize the procedure as follows.  
 To determine its value, we first select a $\kappa_{6}$ value that ranges from -1 to 4, which is slightly wider than the range of the $\kappa$ values determined in the meson cases. Once $\kappa_{6}$ value is chosen, the value of the other parameters $s_{0}$ and $\kappa_{8}$ can be determined by requiring the Borel curve for the mass  to be flattest and stable. Then we  change the $\kappa_{6}$ value  and repeat the procedure and compare the  Borel mass curves.  The final values of $\kappa_{6}$ and $\kappa_8$ are determined from choosing the most stable Borel mass curve among the final Borel mass curves.  It should be noted that although we set the $\kappa_{6}$ range, 
 the stable baryonic sum rule for the mass occurs only for a limited value of $\kappa$'s such that as we approach the lower and upper limit of the range that we chose for $\kappa_6$, the sum rule become far from being stable.  In Table~\ref{baryon-tab}, we summarize the determined $\kappa$ values and masses of nucleon and $\Delta$ in the chiral symmetry restored vacuum.

\begin{center}
\begin{table}[htbp]
\begin{tabular}{c | c c c c c } \hline
	\hline
Particle & $\kappa_6(\sqrt{s_0}(\gev))$ & $\kappa_{8}$ & $S_6/B_6$ &$S_6/B_8$ & $\bar{m}_{sym}$(MeV)\\
\hline
$N$      & 0.05(1.24)& 0.25& -0.95 & -0.75 & 525\\
$\Delta$ & 0.26(1.56)& 0.4 & -0.74 & -0.6 &600\\ \hline
	\hline 
\end{tabular}
\caption{The $\kappa$s are evaluated using sum rules using  the physical values given in Table~\ref{tab1} and the threshold parameter $s_{0}$ in the brackets. $S/B$ indicates the fractional contribution  of the chiral symmetric part to the contribution of the chiral symmetry breaking part for each hadron. $\bar{m}_{sym}$ is the hadronic mass in the chiral symmetry restored vacuum.}
\label{baryon-tab}
\end{table}
\end{center}

\subsection{Nucleon}
\begin{figure}[h]
\centerline{
\includegraphics[width=9 cm]{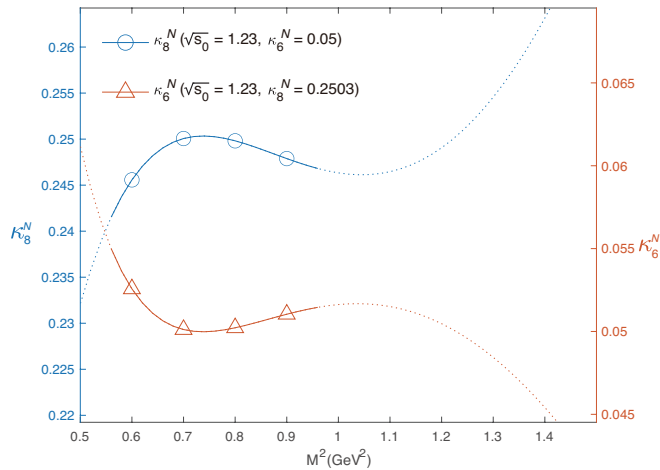}}
\caption{Borel curves for nucleon $\kappa_{6}$(triangle and right y-axis) and $\kappa_{8}$(circle and left y-axis). The curve for one of $\kappa$'s is drawn with the other $\kappa$ taken to be the extremum value inside the Borel window. Unit of $\sqrt{s_{0}}$ is in GeV}
\label{nuclkp}
\end{figure}

\begin{figure}[h]
\centerline{
\includegraphics[width=9 cm]{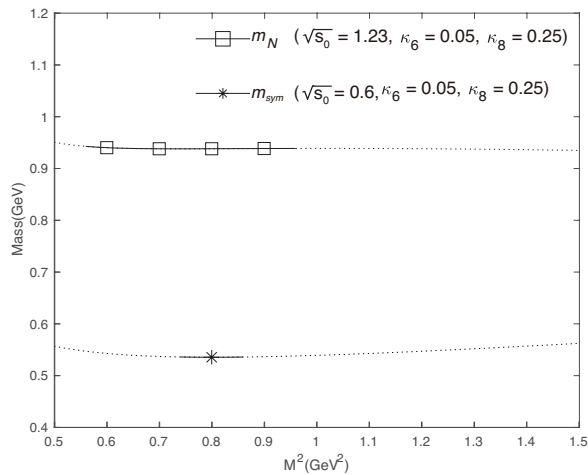}}
\caption{Borel curves for nucleon mass in chiral symmetry broken(square) and symmetry restored(star) vacuum.}
\label{nuclmass}
\end{figure}

We use the sum rule for nucleon constructed from the Ioffe current, known as the optimal choice\cite{Ioffe:1983ju}. Parametrizing the  dimension-6 and dimension-8 operators, respectively, with $\kappa_6^N$ and $\kappa_8^N$,
\begin{equation} 
\begin{split} 
\mathcal{M}^{N}_{6} =&  \kappa_{6}^{N}\frac{2}{3}\ev{\bar{u}u}^{2} = \ev{B_{6}^{N}}_{B} + \ev{S_{6}^{N}}_{S},\\
\mathcal{M}^{N}_{8} =& -\kappa_{8}^{N}\frac{m_{0}^{2}}{6M^{2}}\ev{\bar{u}u}^{2}= \ev{B_{8}^{N}}_{B} + \ev{S_{8}^{N}}_{S},
\end{split} 
\end{equation}
where the $\mathcal{M}$'s are defined in the Appendix, and $B^{N}_{n}$($S^{N}_{n}$) is the dimension-$n$ chiral symmetry breaking(symmetric) operator and given explicitly in Eq.~\ref{bary_dim6} and Eq.~\ref{bary_dim8}.
Then, one can write down the Borel-transformed sum rule as follows.
\begin{equation} 
\begin{split}  \lambda^{2}_{N}\exp(-\frac{m_{N}^{2}}{M^{2}})=& \frac{1}{32\pi^{4}}\bigg(1+\frac{\als}{\pi}\bigg[\frac{53}{12} + \gm[E] \bigg] \bigg)M^{6}E_{2}L^{-4/9}
\\&+\frac{1}{32\pi^{2}}\ev{\frac{\als}{\pi}G^{2}}M^{2}E_{0}L^{-4/9}\\
& +\kappa_{6}^{N}\frac{2}{3}\ev{\bar{u}u}^{2}L^{4/9}-\kappa_{8}^{N}\frac{1}{6}\ev{\bar{u}u}^{2}\frac{m^{2}_{0}}{M^{2}}, \\
\label{nucl_sr}
\end{split} 
\end{equation}
where 
\begin{equation} 
\begin{split} 
L =& \frac{\ln(M/\Lambda_{\mathrm{qcd}})}{\ln(\mu/\Lambda_{\mathrm{qcd}})},\\
m_{0}^{2} =& g_{s}\ev{\bar{u} \sigma_{\mu \nu} G^{\mu \nu}_{a} \lambda^{a} u }/\ev{\bar{u}u},\\
E_{n} = &1-e^{s_{0}/M^{2}}\sum_{j=0}^{n}\frac{1}{j!}\bigg(\frac{s_{0}}{M^{2}}\bigg)^{j},
\end{split} 
\end{equation}
and $\gm[E]$ is the Euler-Mascheroni constant.  
We now use Eq. \eqref{basic-sumrule} using the OPE for the nucleon.  The $\kappa$ values are determined using the procedure discussed in the previous subsection.  To asses the validity of the $\kappa$ values independently, we plot the Borel cures for $\kappa_8$ using Eq. \eqref{basic-sumrule} with determined $\kappa_6^N=0.0425$ value.  The resulting Borel curve is given in  Fig~\ref{nuclkp}.  The figure also shows the Borel curve for $\kappa_6$ with the best fit $\kappa_8^N=0.1546$ value.  
Finally,  as shown in Fig.~\ref{nuclmass}, with the $\kappa$ values, the Borel curve for nucleon mass(square) is stable and well reproduces the physical mass.   

  Here, to compensate for the absence of the sum rule for the chiral partner, we use  the vacuum saturation hypothesis to estimate the value of the  chiral symmetry breaking contribution. Hence, the difference between the total quark operators and the corresponding  chiral symmetry breaking contribution can be identified as the contributions from the chiral symmetric operators. Then the symmetric dimension-6 and dimension-8 contributions are given, respectively, as below.
%
%
%


\begin{equation} 
\begin{split} 
\ev{S_{6}^{N}}_{S} =& (\kappa_{6}^{N}-1) \frac{2}{3}\ev{\bar{u}u}^{2} L^{4/9},\\
\ev{S_{8}^{N}}_{S} =& -(\kappa_{8}^{N}-1) \frac{1}{6}\ev{\bar{u}u}^{2} \frac{m_{0}^{2}}{M^{2}},
\label{nucl_sym}
\end{split} 
\end{equation}
where $\ev{S_{n}^{N}}_{S}$ is the contribution from the chiral symmetric part of the dimension-$n$ operators with $\kappa=1$ in the vacuum saturation hypothesis. The matrix form of the dimension-6 and dimension-8 operators are given in Eq.~(\ref{bary_dim6}). 

The nucleon mass  $m^{N}_{sym}$  in the chiral symmetric vacuum can now be estimated by taking the breaking contributions to be zero while keeping the symmetric operators to their vacuum value.   As can be seen in lower  Borel curve in Fig. \ref{nuclmass},  the mass of the  nucleon and $N^{*}(1535)$ $m^{N}_{sym}$ converge to 500-550 MeV when the symmetry is totally restored. The Borel curve with a star marker in Fig.~\ref{nuclmass} describes the mass, which is also stable.

\subsection{$\Delta$ isobar}

As mentioned in Ref. \cite{PhysRevC.51.2260, Marques:2018mic}, the current $\eta$ for $\Delta^{++}$ given in Table~\ref{tab1} is a unique choice at lowest dimension but  couples to both the spin 3/2 and 1/2 states. So to extract the spin $3/2$ state, we concentrate on the $g_{\mu \nu}$ tensor structure, which has no contribution from the spin $1/2$ state. Introducing $\kappa$s as before,
\begin{equation} 
\begin{split} 
\mathcal{M}^{\Delta}_{8}=& \kappa_{6}^{\Delta}\frac{4}{3}\ev{\bar{u}u}^{2} = \ev{B_{6}^{\Delta}}_{B} + \ev{S_{6}^{\Delta}}_{S},\\
\mathcal{M}^{\Delta}_{8} =& \kappa_{8}^{\Delta}\frac{13}{18M^{2}}m_{0}^{2}\ev{\bar{u}u}^{2} = \ev{B_{8}^{\Delta}}_{B} + \ev{S_{8}^{\Delta}}_{S},
\end{split} 
\end{equation}
where the $\mathcal{M}$'s are defined in the Appendix, and $B^{N}_{n}$($S^{N}_{n}$) is the dimension-$n$ chiral symmetry breaking(symmetric) operator and given explicitly in Eq.~(\ref{bary_dim6}) and Eq.~(\ref{bary_dim8})
Then, the sum rule can be rewritten as  follows.
\begin{equation} 
\begin{split} 
 F_{\Delta} &I^{(0)}_{\Delta}(m_{\Delta}, \Gamma_{\Delta}, M^{2}) = \\
&-\frac{1}{80\pi^{4}}M^{6}E_{2}L^{4/27}+ \frac{5}{288\pi^{2}}M^{2}E_{0}\ev{\frac{\als}{\pi}G^{2}} \\
&- \kappa_{6}^{\Delta}\frac{4}{3}\ev{\bar{u}u}^{2}L^{28/27}+\kappa_{8}^{\Delta}\frac{13}{18M^{2}}m_{0}^{2}\ev{\bar{u}u}^{2}L^{16/27} ,
\label{delsr}
\end{split} 
\end{equation}
where 
\begin{equation} 
\begin{split} 
I_{\Delta}^{(n)} = \int^{s_{0}}_{4m_{\pi}^{2}}e^{-s/M^{2}}\rho^{\mathrm{pole}}(m_{\Delta}, \Gamma_{\Delta},s)s^{n}ds .
\end{split} 
\end{equation}
\begin{figure}[h]
\centerline{
\includegraphics[width=9 cm]{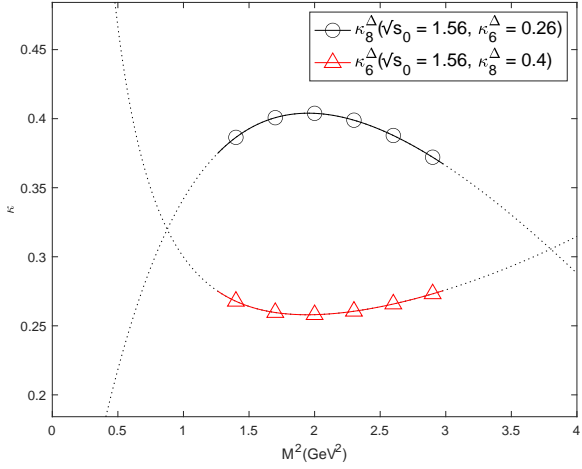}}
\caption{Borel curves for  $\Delta$ $\kappa_{6}$(triangle) and $\kappa_{8}$(circle). The curve for one of $\kappa$s is drawn with the other $\kappa$ taken to be the extremum value. Unit of $\sqrt{s_{0}}$ is in GeV}
\label{delkp}
\end{figure}
\begin{figure}[h]
\centerline{
\includegraphics[width=9 cm]{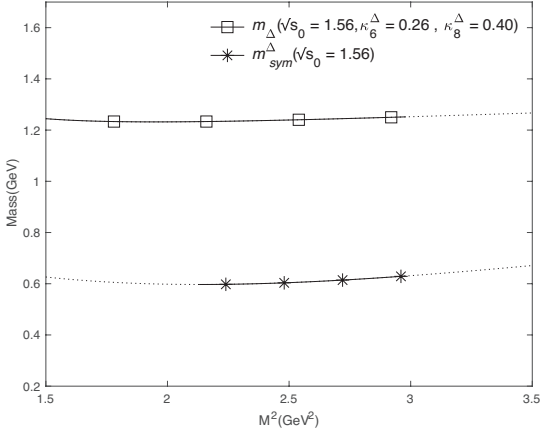}}
\caption{Borel curves for $\Delta$ mass in the chiral symmetry broken(square) and symmetry restored(diamond and star) vacuum. the curves with star markers is the expected $\Delta$ mass. Unit of $\sqrt{s_0}$ is in GeV.}
\label{delmass}
\end{figure}
We first construct the sum rule to estimate the values of the dimension-6 and dimension-8 operators parametrized by $\kappa$'s as before.  As shown in Fig~\ref{delkp}, the Borel curves for $\kappa_{6}$(triangle) and $\kappa_{8}$(open circle) has an extremum value at the same point. 
We can again show that the determined values appropriately reproduces the $\Delta$ mass in the Borel sum rule. Using the $\kappa_6^\Delta=0.26$ and $\kappa_8^\Delta=0.4$  one finds a stable Borel curve for the mass as given in the lines marked by squares in Fig.~\ref{delmass}.  

As in the nucleon case, using the extracted values dimension-6 and dimension-8 quark operators, one can estimate the values for the corresponding chiral symmetric operators appearing in the  $\Delta$ sum rule. 

\begin{equation} 
\begin{split} 
\ev{S_{6}^{\Delta}}_{S} =& -(\kappa_{6}^{\Delta} - 1)\frac{4}{3}\ev{\bar{u}u}^{2}L^{28/27},\\
\ev{S_{8}^{\Delta}}_{S} =& (\kappa_{8}^{\Delta} - 1)\frac{13}{18 M^{2}}m_{0}^{2}\ev{\bar{u}u}^{2}L^{16/27}.
\end{split} 
\end{equation}
By keeping the value of the chiral symmetric operator to be the same as in the vacuum and taking the breaking contribution to be zero, one can extract the $\Delta$ mass in the chiral symmetry restored vacuum $m_{sym}^{\Delta}$ from the Borel  sum rule for the mass. One finds that the mass of $\Delta$ in the symmetry restored vacuum is about 600 MeV.

\section{Summary}

We have studied the masses of hadrons in the chiral symmetry restored vacuum.  This is accomplished through first dividing the quark-operators into chiral symmetry breaking and symmetric operator depending on the contributions from the quark zero modes, which according to Casher-Banks formula is responsible for spontaneous chiral symmetry breaking.  We then estimated the values of these operators separately by studying the sum rules between chiral partners, which have the same contribution from the chiral symmetric operators but where the difference is proportional to chiral symmetry breaking operators.  We have further estimated the contributions of disconnected quark operators by studying the sum rule for $\omega$ and $f_1$ mesons.  
We find that the magnitude of chiral symmetry operators are as big as the chiral symmetry breaking operators.  We further find that the value of chiral symmetry breaking operators are close to the values obtained in the vacuum saturation hypothesis, which automatically relates them to the quark condensate.  On the other hand the value of the chiral symmetric operators could be as large as the breaking operator, which explains the need of effective $\kappa$ parameters to modify the values of the four-quark operators estimated within the vacuum saturation hypothesis to that necessary to reproduce the meson masses in the QCD sum rule approach.  
We then calculated the masses of the hadrons in the chiral symmetry restored vacuum using the QCD sum rule approach by taking the chiral symmetry breaking operators to zero while keeping the values of the other quark operators to their vacuum values. 

We also applied our method to the baryon sector.  In this case, since the chiral partners contribute in the same sum rule, we estimated the values of the dimension-6 and dimension-8 quark operators which contain both the chiral symmetric and breaking operators.  We then used the vacuum saturation hypothesis to estimate the values of the chiral symmetry breaking operators from which we were able to estimate the values for the chiral symmetric operators for both the dimension-6 and dimension-8 operators.  
Again, taking the chiral symmetry breaking operators to zero, we find that the values of the chiral symmetric nucleon and delta masses to be around  500 MeV and 600 MeV, respectively.  

We therefore conclude while chiral symmetry breaking is responsible for the mass difference between chiral partners, a large fraction of the common masses have other non-perturbative origins. 
We further emphasize that our findings show that the vacuum saturation hypothesis that relates quark operators to quark condensate only works for the chiral symmetry breaking operators, which are chiral order parameters as are the quark condensate.

We have shown that masses of all hadron  studied in this work tend to decrease and  the masses of chiral partners become degenerate when the chiral symmetry is restored. Thus the result explicitly demonstrates that the spontaneous breaking of chiral symmetry responsible for the  mass splitting of chiral partners, but is also partly responsible for common  hadronic mass of the chiral partners. Despite their strangeness, $K_1$ and $K^*$ have similar mass as that of $\rho$ and $a_{1}$ in the chiral symmetry restored vacuum.  Meanwhile, the mass difference between $\omega$ and $f_{1}$ is found to remain large even when chiral symmetry is restored. Both the nucleon and the $\Delta$ mass will reduce to about 500 and 600 MeV, respectively, in the chiral symmetry restored vacuum.  
\section*{Acknowledgements}

This work was supported by  by Samsung Science and Technology Foundation under Project Number SSTF-BA1901-04. 


\section*{Appendix}

\subsection{Four-quark operators}
In general, a four-quark operator can be written as follows. 
\begin{equation} 
\begin{split} 
\ev{(\bar{q} \Gamma_{a} q)(\bar{q} \Gamma_{b}q)} =&  \ev{ {\rm Tr} [S \Gamma_a ] {\rm Tr} [S  \Gamma_{b} ]}- \ev{ {\rm Tr} [S \Gamma_{a} S \Gamma_{b} ] ] },
\label{four-q}
\end{split} 
\end{equation}
where $\Gamma$  are  combinations of gamma, color and flavor matrices,  and  $S$ is the quark propagator.  The first term in Eq.~\eqref{four-q} is referred to as the disconnected contribution while the second as the connected. 

First, it is easy to check if the disconnected term is chiral symmetric or breaking operator.  This is accomplished by making use of the first line in Eq.~\eqref{CB-formula} and using the following.
\begin{equation} 
\begin{split} 
{\rm Tr} [S \Gamma_a ] =&  \frac{1}{2} \bigg( {\rm Tr} [(S+i \gamma_5 S i \gamma_5)  \Gamma_a ]+  {\rm Tr} [(S- i \gamma_5 S i \gamma_5)  \Gamma_a ] \bigg).
\label{four-diss}
\end{split} 
\end{equation}
The first term is chiral symmetric and the second breaking term. In terms of the Dirac zero modes, only the second term is proportional to the them.  However, one of them is zero depending on  $\Gamma_a$.  For example, if $\Gamma_a=\gamma_\mu$, the breaking term is zero so that the total disconnect contribution is chiral symmetric.  The term is automatically zero if $\Gamma_a$ includes a flavor matrix. 

Second, as for the connected contribution, one can decompose the matrix into chiral symmetric and breaking part. If $\Gamma_a$ does not include a flavour matrix, one can subtract and add a similar matrix with the flavour matrix inserted so as to isolate the connected contribution. 
\begin{equation} 
\begin{split} 
\ev{(\bar{q} \Gamma_{a} q)(\bar{q} \Gamma_{b}q)} =& \bigg( \ev{(\bar{q} \Gamma_{a}  q)(\bar{q} \Gamma_{b}q)} -\ev{(\bar{q} \Gamma_{a} \tau_i q)(\bar{q} \Gamma_{b} \tau_i q)} \bigg) \\
& + \ev{(\bar{q} \Gamma_{a} \tau_i q)(\bar{q} \Gamma_{b} \tau_i q)} ,
\label{four-tau}
\end{split} 
\end{equation}
where the first line contains only disconnected contribution while the second line only connected.   Therefore, one is left with determining the chiral property of the connected diagrams.  This is accomplished by the following separation.
\begin{widetext}
\begin{equation} 
\begin{split} 
\ev{(\bar{q} \Gamma_{a} \tau_i q)(\bar{q} \Gamma_{b} \tau_i q)} =& \frac{1}{2} \bigg( \ev{(\bar{q} \Gamma_{a} \tau_i  q)(\bar{q} \Gamma_{b}q)} -\ev{(\bar{q} \Gamma_{a} \tau_i i\gamma_5 q)(\bar{q} i\gamma_5 \Gamma_{b} \tau_i q)} \bigg)  + \frac{1}{2} \bigg( \ev{ (\bar{q} \Gamma_{a} \tau_i  q)(\bar{q} \Gamma_{b}q)} +\ev{(\bar{q} \Gamma_{a} \tau_i i\gamma_5 q)(\bar{q} i\gamma_5 \Gamma_{b} \tau_i q)} \bigg) \\
 =& \frac{1}{2} \bigg( \ev{ {\rm Tr} [ S \Gamma_{a} \tau_i (S-i\gamma_5 S i \gamma_5) \Gamma_b \tau_i ] } \bigg)
 +
 \frac{1}{2} \bigg( \ev{ {\rm Tr} [ S \Gamma_{a} \tau_i (S+i\gamma_5 S i \gamma_5) \Gamma_b \tau_i ] } \bigg)
  \\
=& \ev{(\bar{q} \Gamma_{a} \tau_i q)(\bar{q} \Gamma_{b} \tau_i q)}_B+\ev{(\bar{q} \Gamma_{a} \tau_i q)(\bar{q} \Gamma_{b} \tau_i q)}_S ,
\label{four-con}
\end{split} 
\end{equation}
\end{widetext}
where the subscript $B$ and $S$ represent the chiral symmetry breaking and symmetric parts respectively.

\subsection{Fierz Transformation}
Every four-quark operator that appears in baryonic OPE sides has the following form\cite{Koike:1993sq}. 
\begin{equation} 
\begin{split} 
\ev{\mathcal{O}_{6}} =& \frac{2}{3}\ev{(\bar{q}_{1}\Gamma_{o} q_{2})(\bar{q}_{3} \Gamma_{p}q_{4})} \\
&- \frac{1}{2}\ev{(\bar{q}_{1}\Gamma_{o}\lambda_{a}q_{2})(\bar{q}_{3}\Gamma_{p}\lambda^{a}q_{4})},
\label{bary_fq}
\end{split} 
\end{equation}
where $\Gamma_{\alpha}$ and  $\lambda^{a}$ are a Dirac matrix and a Gell-Mann matrix, respectively. Using the Fierz completeness relation for the color $SU(3)$, 
\begin{equation} 
\begin{split} 
\mathds{1}_{ij}\mathds{1}_{kl} = \frac{1}{3}\mathds{1}_{il}\mathds{1}_{kj} + \sum_{A=1}^{8}\frac{1}{2}\lambda^{A}_{il}\lambda^{A}_{kj},
\label{col_fi}
\end{split} 
\end{equation}
the Gell-Mann matrices in the operator can be eliminated. 
\begin{eqnarray} 
\langle(\bar{q}_{1}\Gamma_{o}\lambda_{a}q_{2})&(\bar{q}_{3}&\Gamma_{p}\lambda^{a}q_{4})\rangle =\nonumber\\
&-\frac{2}{3}&\ev{(\bar{q}_{1}\Gamma_{o}q_{2})(\bar{q}_{3}\Gamma_{p}q)} \nonumber\\
&-\frac{1}{8}&\Tr[\Gamma_{o}\Gamma^{r}\Gamma_{p}\Gamma^{s}]\ev{(\bar{q}_{1}\Gamma_{r}q_{4})(\bar{q}_{3}\Gamma_{s}q_{2})},
\end{eqnarray}
where the Dirac and Gell-Mann matrices are normalized as $\Tr[\Gamma_{r}\Gamma^{s}] = 4 \delta_{r}^{s}$ and $\Tr[\lambda^{a}\lambda_{b}] = 2 \delta^{a}_{b}$, respectively. Applying the transformation reduces the number of four-quark operators in the baryonic OPE sides by half. 
When it comes to the pure-flavored four-quark operators, the transformation can be represented linearly. 
\begin{widetext}
\begin{equation} 
\begin{split} 
\begin{pmatrix}
  \ev{(\bar{u}\lambda^{a}u)^{2}}	 \\
  \ev{(\bar{u}\gm[5]\lambda^{a}u)^{2}}\\
  \ev{(\bar{u}\gm \lambda^{a}u)^{2}}\\
  \ev{(\bar{u}\gm[5]\gm \ld u)^{2}}\\
  \ev{(\bar{u}\sigma_{\mu \nu}\ld u)^{2}}
  \end{pmatrix}
  = 
  \begin{pmatrix}
  	-7/6 & -1/2 &-1/2 &  1/2 & -1/4 \\
    -1/2 & -7/6 & 1/2 & -1/2 & -1/4 \\
      -2 &    2 & 1/3 &    1 &    0 \\
       2 &   -2 &   1 &  1/3 &    0 \\
      -6 &   -6 &   0 &    0 &  1/3
  \end{pmatrix}
  \begin{pmatrix}
  	\ev{(\bar{u}u)^{2}} \\
  	\ev{(\bar{u}\gm[5]u)^{2}} \\
  	\ev{(\bar{u}\gm u)^{2}}\\
  	\ev{(\bar{u}\gm[5] \gm u)^{2}}\\
  	\ev{(\bar{u}\sigma_{\mu \nu} u)^{2}}
  \end{pmatrix}.
\end{split} 
\end{equation}
\end{widetext}

\subsection{Separation of breaking and symmetric operators}

Using the prescription discussed in Appendix A, one can categorize all the quark operators appearing in the nucleon sum rule as follows.
With the following identity,
\begin{equation} 
\begin{split} 
 \langle(\bar{u}\Gamma_{1} d)(\bar{d} \Gamma^{1} u)\rangle =\ev{(\bar{u}\Gamma_{1}u)(\bar{u}\Gamma^{1}u)} - \ev{(\bar{u}\Gamma_{1} u)(\bar{d} \Gamma^{1} d)},
 \label{pmm}
\end{split} 
\end{equation}
there are 5(Dirac structure) $\times$ 2(pure or mixed flavor) types of four-quark condensates in the baryonic OPE sides. These four-quark condensates also can be divided into chiral symmetry breaking and symmetric condensates according to the contribution of the zero modes. The operators denoted with subscript $S$ ($\ev{...}_{S}$) and $B$ ($\ev{...}_{B}$) denote chiral symmetric and breaking operators, respectively. 
\begin{equation} 
\begin{split} 
 1. \ev{(\bar{u}u)(\bar{u}u)} =& \ev{(\bar{u}u)(\bar{u}u)- (\bar{u}d)(\bar{d}u)}_{B} \\
 &+ \ev{(\bar{u}d)(\bar{d}u)}\\
 =&\ev{(\bar{u}u)(\bar{d}d)}_{B} \\
 &+ \frac{1}{2}\ev{(\bar{u}d)(\bar{d}u)+(\bar{u}\gm[5] d)(\bar{d} \gm[5] u)}_{B}\\
  &+ \frac{1}{2}\ev{(\bar{u}d)(\bar{d}u)-(\bar{u}\gm[5] d)(\bar{d} \gm[5] u)}_{S}.
  \label{scalar-sep}
\end{split} 
\end{equation}
Since the pure flavored condensates contain connected and disconnected pieces, we first divide the condensate into the two. Then, the connected piece is separated into the breaking and symmetric condensates. In the same way, the other remaining operators can be rewritten as the following.
\begin{widetext} 
\begin{equation} 
\begin{split} 
&2. \ev{(\bar{u} \gm[5] u)(\bar{u} \gm[5] u)} = \ev{(\bar{u}\gm[5] u)(\bar{d} \gm[5] d)}_{B}+\frac{1}{2}\ev{(\bar{u}d)(\bar{d}u)+(\bar{u}\gm[5] d)(\bar{d} \gm[5] u)}_{B}-\frac{1}{2}\ev{(\bar{u}d)(\bar{d}u)-\ev{\bar{u}\gm[5] d}(\bar{d} \gm[5] u)}_{S},\\
&3. \ev{(\bar{u}\gm u)(\bar{u} \gm u)} =  \ev{(\bar{u}\gm u)(\bar{d} \gm d)}_{S}+\frac{1}{2}\ev{(\bar{u}\gm d)(\bar{d} \gm u)+(\bar{u} \gm[5] \gm d)(\bar{d} \gm[5] \gm u)}_{S}\\
&\qquad \qquad \qquad \qquad\qquad+\frac{1}{2}\ev{(\bar{u}\gm d)(\bar{d} \gm u)-(\bar{u} \gm[5] \gm d)(\bar{d} \gm[5] \gm u)}_{B},\\
&4. \ev{(\bar{u}\gm[5]\gm u)(\bar{u} \gm[5]\gm u)} =  \ev{(\bar{u}\gm[5]\gm u)(\bar{d}\gm[5] \gm d)}_{S}+\frac{1}{2}\ev{(\bar{u}\gm d)(\bar{d} \gm u)+(\bar{u} \gm[5] \gm d)(\bar{d} \gm[5] \gm u)}_{S}\\
&\qquad \qquad \qquad \qquad\qquad-\frac{1}{2}\ev{(\bar{u}\gm d)(\bar{d} \gm u)-(\bar{u} \gm[5] \gm d)(\bar{d} \gm[5] \gm u)}_{B},\\
&5. \ev{(\bar{u}\sigma_{\mu \nu} u)(\bar{u} \sigma_{\mu \nu} u)} =  \ev{(\bar{u}\sigma_{\mu \nu} u)(\bar{d} \sigma_{\mu \nu} d)}_{B} +\ev{(\bar{u}\sigma_{\mu \nu} d)(\bar{d}\sigma_{\mu \nu}u)}_{B}.
\label{rest-sep}
\end{split} 
\end{equation}
\end{widetext}

For mixed flavor operators, only disconnected contributions exist so that the chiral symmetry properties can be determined as discussed in Appendix A.
Note that since the color matrices $\ld$ have no impact on the separation rules, the same formula as given in Eq. \eqref{rest-sep} applies when the quark anti-quark pair contains a Gell-Mann matrix with the index contracted with that in the other pair.

\subsection{Baryon quark operators}

In this subsection, we summarize the quark operators separated into chiral symmetric and breaking parts in the Baryon sum rules.
  The dimension-6 four quark operators appearing in the baryonic OPE sides are given, explicitly.

\begin{equation} 
\begin{split} 
\mathcal{M}_{6}^{N} = & \ev{B_{6}^{N}}_{B} + \ev{S_{6}^{N}}_{S},\\
\mathcal{M}_{6}^{\Delta} =& \ev{B_{6}^{\Delta}}_{B} + \ev{S_{6}^{\Delta}}_{S},
\label{sep-nd}
\end{split} 
\end{equation}
where
\begin{widetext}
\begin{equation} 
\begin{split} 
\ev{B^{N}_{6}}_{B} =& -\ev{(\bar{u}u)(\bar{d}d)}_{B} + \ev{(\bar{u}\gm[5] u)(\bar{d} \gm[5] d)}_{B} - \frac{1}{2}\ev{(\bar{u}\gm d)(\bar{d} \gm u) - (\bar{u}\gm \gm[5]d)(\bar{d} \gm \gm[5] u)}_{B},\\
\ev{S^{N}_{6}}_{S} =&  -3\ev{(\bar{u}\gm u)(\bar{d}\gm d)}_{S} - \ev{(\bar{u}\gm \gm[5]d)(\bar{d} \gm \gm[5] u)}_{S} +2 \ev{(\bar{u}\gm d)(\bar{d} \gm u) + (\bar{u}\gm \gm[5]d)(\bar{d} \gm \gm[5] u)}_{S},\\
\ev{B^{\Delta}_{6}}_{B} =& -2\ev{(\bar{u}u)(\bar{d}d)}_{B}+2\ev{(\bar{u}\gm[5] u)(\bar{d}\gm[5]d)}_{B}-\ev{(\bar{u} \gm d)(\bar{d} \gm u)-(\bar{u} \gm \gm[5] d)(\bar{d}\gm \gm[5] u)}_{B},\\
\ev{S^{\Delta}_{6}}_{S}=&-2\ev{(\bar{u}d)(\bar{d}u)-(\bar{u}\gm[5]d)(\bar{d} \gm[5] u)}_{S} -\ev{(\bar{u}\gm u)(\bar{d}\gm d)}_{S}+\ev{(\bar{u} \gm \gm[5] u)(\bar{d} \gm \gm[5] d)}_{S},
\label{bary_dim6}
\end{split} 
\end{equation}
\end{widetext}
 $\mathcal{M}_{N}^{6}$ and $\mathcal{M}_{\Delta}^{6}$ are the dimension-6 terms that appear in OPE side of nucleon and $\Delta$, respectively. For simplicity, the anomalous dimensional correction factors are omitted.

For notational convenience, the followings are introduced.
\begin{widetext}
\begin{equation} 
\begin{split} 
D\{\ev{(\bar{q}_{1} \Gamma_{a} q_{2})(\bar{q}_{3} \Gamma_{b} q_{4})}\} \equiv& \frac{1}{2} \bigg( \ev{(\bar{q}_{1} \Gamma_{a} D^{2} q_{2})(\bar{q}_{3} \Gamma_{b} q_{4})} + \ev{(\bar{q}_{1} \Gamma_{a} q_{2})(\bar{q}_{3} \Gamma_{b} D^{2}q_{4})}\bigg),\\
G\{\ev{(\bar{q}_{1} \Gamma_{a} q_{2})(\bar{q}_{3} \Gamma_{b} q_{4})}\} \equiv& \frac{1}{2} \bigg( \ev{(\bar{q}_{1} \Gamma_{a}\sigma \cdot G q_{2})(\bar{q}_{3} \Gamma_{b} q_{4})} + \ev{(\bar{q}_{1} \Gamma_{a} q_{2})(\bar{q}_{3} \Gamma_{b} \sigma \cdot G q_{4})}\\
&+\ev{(\bar{q}_{1} \Gamma_{a}\sigma \cdot G^{A} q_{2})(\bar{q}_{3} \Gamma_{b}\lambda_{A} q_{4})} +\ev{(\bar{q}_{1} \Gamma_{a}\lambda_{A} q_{2})(\bar{q}_{3} \Gamma_{b} \sigma \cdot G^{A} q_{4})} \bigg),
\label{not-dg}
\end{split} 
\end{equation}
\end{widetext}
where 
\begin{eqnarray} 
D^{2} &=& D_{\mu}D^{\mu},\nonumber\\
G_{\mu \nu}& =& G_{\mu \nu}^{A} \lambda_{A},\nonumber\\
(\sigma\cdot G)^{ab}_{ij} &=& (\sigma_{\mu \nu})_{ij} G^{\mu \nu}_{A} (\lambda^{A})^{ab},\nonumber\\
\sigma_{\mu \nu} &=& \frac{i}{2}[\gamma_{\mu},\gamma_{\nu}].
\end{eqnarray}
Since gluon field strength tensor and gauge covariant derivative operators do not affect separation rules, the dimension-8 operators in right-hand side of Eq~(\ref{not-dg}) have the same chiral separation property as the dimension-6 operator inside the  curly bracket. Namely 
\begin{widetext}
\begin{equation} 
\begin{split} 
D\{\ev{(\bar{q}_{1} \Gamma_{a} q_{2})(\bar{q}_{3} \Gamma_{b} q_{4})}_{B,S}\} =& \frac{1}{2} \bigg( \ev{(\bar{q}_{1} \Gamma_{a} D^{2} q_{2})(\bar{q}_{3} \Gamma_{b} q_{4})}_{B,S} + \ev{(\bar{q}_{1} \Gamma_{a} q_{2})(\bar{q}_{3} \Gamma_{b} D^{2}q_{4})}_{B,S}\bigg),\\
G\{\ev{(\bar{q}_{1} \Gamma_{a} q_{2})(\bar{q}_{3} \Gamma_{b} q_{4})}_{B,S}\} =& \frac{1}{2} \bigg( \ev{(\bar{q}_{1} \Gamma_{a}\sigma \cdot G q_{2})(\bar{q}_{3} \Gamma_{b} q_{4})}_{B,S} + \ev{(\bar{q}_{1} \Gamma_{a} q_{2})(\bar{q}_{3} \Gamma_{b} \sigma \cdot G q_{4})}_{B,S}
\\&+\ev{(\bar{q}_{1} \Gamma_{a}\sigma \cdot G^{A} q_{2})(\bar{q}_{3} \Gamma_{b}\lambda_{A} q_{4})}_{B,S} + \ev{(\bar{q}_{1} \Gamma_{a}\lambda_{A} q_{2})(\bar{q}_{3} \Gamma_{b} \sigma \cdot G^{A} q_{4})}_{B,S} \bigg).
\end{split} 
\end{equation}
\end{widetext}
The dimension-8 operators appearing in nucleon and $\Delta$ OPE side are given, respectively, as
\begin{equation} 
\begin{split} 
\mathcal{M}_{8}^{N} =& \ev{B_{8}^{N}}_{B} + \ev{S_{8}^{N}}_{S}, \\
\mathcal{M}_{8}^{\Delta} =& \ev{B_{8}^{\Delta}}_{B} + \ev{S_{8}^{\Delta}}_{S}.
\label{dim8quark}
\end{split} 
\end{equation}
where
\begin{widetext}
\begin{equation} 
\begin{split} 
\ev{B_{8}^{N}}_{B} =& -\frac{1}{M^{2}} D\{ B_{6}^{N}\}+\frac{1}{6M^{2}} G\{\ev{(\bar{u}u)(\bar{d}d)}_{B} \} +\frac{1}{12 M^{2}} G\{ \ev{(\bar{u}d)(\bar{d}u) + (\bar{u}\gm[5] d)(\bar{d} \gm[5] u)}_{B} \},\\
\ev{S_{8}^{N}}_{S} =& -\frac{1}{M^{2}} D\{ S_{6}^{N}\} +\frac{1}{12 M^{2}} G\{ \ev{(\bar{u}d)(\bar{d}u) - (\bar{u} \gm[5] d)(\bar{d} \gm[5] u)}_{S}\},\\
\ev{B_{8}^{\Delta}}_{B}=&-\frac{1}{M^{2}} D\{ B_{6}^{\Delta}\}+ \frac{1}{18M^{2}}G\{ \ev{(\bar{u}u)(\bar{d}d)}_{B}  \}  +\frac{1}{36M^{2}} G \{\ev{(\bar{u}d)(\bar{d} u) +(\bar{u} \gm[5] d)(\bar{d} \gm[5] u)}_{B}\}, \\
\ev{S_{8}^{\Delta}}_{S} = & -\frac{1}{M^{2}} D\{ S_{6}^{\Delta}\}+\frac{1}{36M^{2}} G\{ \ev{(\bar{u} d)(\bar{d} u) - (\bar{u} \gm[5] d )(\bar{d}\gm[5] u)}_{S} \}.
\label{bary_dim8}
\end{split} 
\end{equation}
\end{widetext}
For simplicity, the anomalous dimensional correction factors are omitted.


\begin{thebibliography}{99} 

\bibitem{Wilczek:1999be}
F.~Wilczek,
Phys. Today \textbf{52N11}, 11-13 (1999)


\bibitem{Wilczek:2012sb}
F.~Wilczek,
Central Eur. J. Phys. \textbf{10}, 1021-1037 (2012)
[arXiv:1206.7114 [hep-ph]].


\bibitem{Nambu:1961tp}
Y.~Nambu and G.~Jona-Lasinio,
Phys. Rev. \textbf{122}, 345-358 (1961).

\bibitem{Nambu:1961fr}
Y.~Nambu and G.~Jona-Lasinio,
Phys. Rev. \textbf{124}, 246-254 (1961).

\bibitem{Hatsuda:1985eb}
  T.~Hatsuda and T.~Kunihiro,
  Phys.\ Rev.\ Lett.\  {\bf 55}, 158 (1985).

\bibitem{Brown:1991kk}
  G.~E.~Brown and M.~Rho,
  Phys.\ Rev.\ Lett.\  {\bf 66}, 2720 (1991).

\bibitem{Hatsuda:1991ez}
  T.~Hatsuda and S.~H.~Lee,
  Phys.\ Rev.\ C {\bf 46}, no. 1, R34 (1992).


\bibitem{Leupold:2009kz}
  S.~Leupold, V.~Metag and U.~Mosel,
  Int.\ J.\ Mod.\ Phys.\ E {\bf 19}, 147 (2010).


\bibitem{Hayano:2008vn}
For review see, R.~S.~Hayano and T.~Hatsuda,
  Rev.\ Mod.\ Phys.\  {\bf 82}, 2949 (2010).

\bibitem{Ichikawa:2018woh}
  M.~Ichikawa {\it et al.},
  arXiv:1806.10671 [physics.ins-det].


\bibitem{Metag:2017yuh}
  V.~Metag, M.~Nanova and E.~Y.~Paryev,
  Prog.\ Part.\ Nucl.\ Phys.\  {\bf 97}, 199 (2017).


\bibitem{Ohnishi:2019cif}
H.~Ohnishi, F.~Sakuma and T.~Takahashi,
Prog. Part. Nucl. Phys. \textbf{113}, 103773 (2020).

\bibitem{Salabura:2020tou}
P.~Salabura and J.~Stroth,
[arXiv:2005.14589 [nucl-ex]].

\bibitem{Weinberg:1967kj}
  S.~Weinberg,
  Phys.\ Rev.\ Lett.\  {\bf 18}, 507 (1967).
  

\bibitem{Song:2018plu}
T.~Song, T.~Hatsuda and S.~H.~Lee,
Phys. Lett. B \textbf{792}, 160-169 (2019).


\bibitem{Lee:2019tvt}
S.~H.~Lee,
JPS Conf. Proc. \textbf{26}, 011012 (2019)
[arXiv:1904.09064 [nucl-th]].



\bibitem{Kim:2020zae}
J.~Kim and S.~H.~Lee,
Phys. Rev. D \textbf{103}, no.5, L051501 (2021)


\bibitem{BC} T. Banks and A. Casher, Nucl. Phys.  {\bf B 169} 103 (1980).

\bibitem{Shifman:1978bx}
  M.~A.~Shifman, A.~I.~Vainshtein and V.~I.~Zakharov,
  Nucl.\ Phys.\  B {\bf 147}, 385 (1979); 
  M.~A.~Shifman, A.~I.~Vainshtein and V.~I.~Zakharov,
  Nucl.\ Phys.\  B {\bf 147}, 448 (1979).


\bibitem{Leupold:1998bt}
S.~Leupold and U.~Mosel,
Phys. Rev. C \textbf{58}, 2939-2957 (1998)
doi:10.1103/PhysRevC.58.2939
[arXiv:nucl-th/9805024 [nucl-th]].

\bibitem{Leupold:2001hj}
S.~Leupold,
Phys. Rev. C \textbf{64}, 015202 (2001)
doi:10.1103/PhysRevC.64.015202
[arXiv:nucl-th/0101013 [nucl-th]].


\bibitem{Cohen:1996ng}
  T.~D.~Cohen,
  Phys.\ Rev.\ D {\bf 54}, 1867 (1996).

\bibitem{ParticleDataGroup:2020ssz}
P.~A.~Zyla \textit{et al.} [Particle Data Group],
PTEP \textbf{2020}, no.8, 083C01 (2020)
doi:10.1093/ptep/ptaa104

\bibitem{Hatsuda:1992bv}
T.~Hatsuda, Y.~Koike and S.~H.~Lee,
Nucl. Phys. B \textbf{394}, 221-266 (1993)
doi:10.1016/0550-3213(93)90107-Z


\bibitem{Gubler:2016djf}
P.~Gubler, T.~Kunihiro and S.~H.~Lee,
Phys. Lett. B \textbf{767}, 336-340 (2017)


\bibitem{Gubler2013}
P.~Gubler, A Bayesian Analysis of QCD Sum Rules, Springer Theses, Springer Japan, 2013.


\bibitem{Ioffe:1983ju}
B.~L.~Ioffe and A.~V.~Smilga,
Nucl. Phys. B \textbf{232}, 109-142 (1984)


\bibitem{PhysRevC.51.2260}
X.~Jin, Phys. Rev. C. \textbf{51} 2260-2263 (1995)

\bibitem{Marques:2018mic}
J.~Marques L.,~S.~H.~Lee, A.~Park, R.~D.~Matheus and K.~S.~Jeong,
Phys. Rev. C \textbf{98}, no.2, 025206 (2018)


\bibitem{Koike:1993sq}
Y.~Koike,
Phys. Rev. D \textbf{48}, 2313-2323 (1993)

\end{thebibliography}
\end{document}